\documentclass[12pt]{article}
\usepackage{graphicx}

\usepackage{hyperref}
\hypersetup{
    colorlinks=true,
    linkcolor=blue,
    filecolor=magenta,      
    urlcolor=cyan,
}
\usepackage[lofdepth,lotdepth]{subfig}

\newcommand\pubnumber{CMS-CR-2017/404}
\newcommand\pubdate{\today}

\def\institute{Chinese Academy of Sciences\\
Institute of High Energy Physics, Beijing, CHINA}

\def\Title#1{\begin{center} {\Large #1 } \end{center}}
\def\Author#1{\begin{center}{ \sc #1} \end{center}}
\def\Address#1{\begin{center}{ \it #1} \end{center}}

\newcommand\pubblock{\rightline{\begin{tabular}{l} \pubnumber\\
         \pubdate  \end{tabular}}}
\newcommand{\MG} {\textsc{mg5}\_a\textsc{mc@nlo}}
\newcommand{\PYTHIA} {\textsc{pythia}}
\newcommand*{\ttbar}{$t\overline{t}$}
\newcommand*{\alfa}{$\alpha_s^{ISR}$}
\newcommand*{\hdamp}{$h_{damp}$}
\newcommand{\PHEG} {\textsc{powheg v2}}

\newcommand{\CP}{\textsc{cuetp8m2t4}}
\newcommand{\CO}{\textsc{cuetp8m1}}
\newcommand{\HERWIG}{\textsc{herwig++}}
\newcommand{\HERWIGalone}{\textsc{herwig}}
\newcommand{\SHERPA}{\textsc{sherpa}}

\newenvironment{Abstract}{\begin{quotation}  }{\end{quotation}}
\newenvironment{Presented}{\begin{quotation} \begin{center} 
             PRESENTED AT\end{center}\bigskip 
      \begin{center}\begin{large}}{\end{large}\end{center} \end{quotation}}

\def\beq{\begin{equation}}
\def\eeq#1{\label{#1}\end{equation}}
\def\eeqn{\end{equation}}

\def\beqa{\begin{eqnarray}}
\def\eeqa#1{\label{#1}\end{eqnarray}}
\def\eeqan{\end{eqnarray}}

\let\bar=\overbar

\def\Dslash{\not{\hbox{\kern-4pt $D$}}}
\def\dslash{\not{\hbox{\kern-2pt $\del$}}}

\def\msb{{\bar{\ssstyle M \kern -1pt S}}}

\begin{document}
\begin{titlepage}
\pubblock

\vfill
\Title{Top Quark Modeling and Generators in CMS}
\vfill
\Author{Efe Yazgan on behalf of the CMS Collaboration}
\Address{\institute}
\vfill
\begin{Abstract}
Recent top quark event modeling studies done using LHC proton-proton collision data collected with the CMS detector at centre of mass energies of 8 and 13 TeV and state-of-the-art theoretical predictions are summarized.~A new factorized approach for parton shower uncertainties is presented. A~top quark specific \PYTHIA8~CMS tune, along with tunes using new color reconnection models, is discussed.~The possibility of having a consistent choice of parton distribution function in the matrix element and the parton shower is demonstrated with tunes constructed with leading, next-to-leading, and next-to-next-to-leading order versions of NNPDF3.1 set compared to minimum bias and underlying event data. 

\end{Abstract}
\vfill
\begin{Presented}
$9^{th}$ International Workshop on Top Quark Physics\\
Braga, Portugal,  September 17--22, 2017
\end{Presented}
\vfill
\end{titlepage}
\def\thefootnote{\fnsymbol{footnote}}
\setcounter{footnote}{0}

\section{Introduction}
Top quark measurements provide important tests of QCD and of the consistency of the standard model (SM). Better understanding of perturbative and non-perturbative effects is required to obtain the highest possible precision in the measurement of top quark properties, in particular, the top quark mass, and its interpretation.~Differential measurements done with well-defined top-quark objects are also important to improve the accuracy of predictions in different phase space regions in searches for beyond the SM effects.~The uncertainties in the measurements and the predictions need to be at a level where deviations from the predictions from the Monte Carlo (MC) codes or deviations due to new physics effects may become visible.~State-of-the-art  next-to-leading order (NLO) matrix element (ME) event generators interfaced to more recent parton shower (PS) codes used in LHC Run 2 may provide better modeling and eventually reduce the major theoretical uncertainties.~In this note, a selection of recent top quark event modeling studies from the CMS\cite{CMS} Collaboration are discussed. 

\section{Particle Level Top Quark}
Simulations at NLO take the finite width of the top quark into account to model the off-shell production of top quarks and their interference with the backgrounds.~In these calculations the concept of top quark as a particle is not well-defined and has strong dependence on the choice of the MC generator.~One can only use the kinematics of the final-state particles for unambiguous comparisons to theory predictions.~A particle-level top quark (so-called pseudo-top quark) can be constructed from the final-state objects after hadronization. Using particle-level top quarks yields smaller uncertainties from non-perturbative effects and from acceptance corrections thanks to the similar phase definitions at the particle and detector levels, thereby reducing the MC dependence.~The details of particle-level top quark definitions and their adoption in the RIVET \cite{RIVETrep} framework in the official CMS reconstruction code are discussed in \cite{Collaboration:2267573} as a fundamental aspect for current and future measurements of differential production cross sections in both top quark pair (\ttbar) and single-top quark production. 

\section{Factorized PS and Hadronization Uncertainties}
The top quark is a colored particle that decays into another colored particle, the b-quark, and most of the times accompanied by extra jets.~The predictions are reliable only after the ME is interfaced to the parton shower.~In the simulation of top quark events, ambiguities arise from the shower scales, ME-PS matching, soft non-perturbative QCD effects, color reconnection, fragmentation, flavour response and hadronization, and 
semileptonic B hadron branching fraction.~For most measurements, experimental collaborations usually compare predictions from two different PS codes, e.g. \PYTHIA~ vs \HERWIG.~However, in experiments jet energy corrections and b-tagging scale factors are typically derived based on a single PS code. Therefore, comparing two PS codes requires ad-hoc corrections. In some cases, even after corrections large discrepancies remain leading to overestimated or not-so-well understood PS uncertainties.~To get better insights in the PS uncertainties, CMS adopted a new method to calculate PS uncertainties through variations in a single parton shower simulation (\PYTHIA8).~These variations cover uncertainties in the modeling of perturbative and non-perturbative QCD effects in a parton shower MC.~The individual uncertainty sources, their corresponding parameters/quantities, and variations are shown in Table~\ref{tab:unc}.~Information on the possibility to determine the uncertainties using event weights is also provided. 

\begin{table}[!htbp]
\begin{center}
\caption{
Factorized PS and hadronization uncertainties. The uncertainty source, the corresponding quantities, variations and corresponding references are displayed.~Information on the possibility to determine the uncertainties using event weights is also provided. 
}
\label{tab:unc}
\begin{center}
\scalebox{0.7}{
\begin{tabular}{lcccc}
\hline
Source                     & Handle & Weights & Variation & Note/Ref. \\
\hline
Shower scales         &  ISR/FSR            &    No          &    0.5-2.0             &      [1,2] /-            \\ \hline
ME-PS matching        &    hdamp          &   No           &  1.58$^{+0.66}_{-0.59}$m$_{\rm t}$               &    -/\cite{CMS-PAS-TOP-16-021}                    \\ \hline
Soft QCD                &  UE parameters            &   No           &        up/down           &  [3]/\cite{CMS-PAS-TOP-16-021}                    \\ \hline
Color reconnection         & MPI based,         &       No       &     compare models            &    [4]/-                 \\
(Odd clusters)         &     gluon move         &              &                 &                            \\
                               & QCD-inspired &              &                 &                            \\ \hline
Fragmentation        &  momentum transfer from            &              &        Vary   x${\rm _b} $         &           [5]/\cite{CMS-PAS-TOP-16-022}                 \\
	                       & b-quark to B-hadron                    &   Yes     		&  parameter within  & \\ 
	                       &  x${\rm _b}$=p$_{\rm T}$(B)/p$_{\rm T}$(b-jet)        &    &  uncertainties    &   \\ 
	                       &              &              & from LEP/SLD fits  & \\\hline
Flavor response/        &   Pythia vs Herwig           &       No       &     vary JES for each flavour            &         -/-                   \\
hadronization         &              &              &           for light, g, c, b      &                           \\ \hline
Decay tables         &   B semi-leptonic BR           &      Yes        &    +0.77\%/-0.45\%             &        [6]/-                    \\ \hline
\end{tabular}
}
\end{center}

\footnotesize{ 
\raggedright [1] Since \PYTHIA8.230, it is possible to calculate shower scales with event weights.\\
\raggedright [2] FSR variations scaled down by $\sqrt{2}$ based on LEP data.   \\
\raggedright [3] Multiparton Interactions (MPI) or Color Reconnection (CR) strength do not affect resonance decays. \\
\raggedright [4] CR affects resonance decays. \\
\raggedright [5] Re-weight x$_{\rm b}$.  \\
\raggedright [6] Re-weight the fraction of semi-leptonic b jets by the PDG values (scale $\Lambda_{\rm b}$ to match PDG value).  \\
}
\end{center}
\end{table}
\section{A Top Quark Specific Event Tune}
The predictions of the NLO  ME generators + \PYTHIA ~\CO tune (based on the Monash tune \cite{Skands:2014pea}) overshoot the 
CMS data for large jet multiplicities, while all other distributions are modelled well (except the transverse momentum ($p_T$) of the top quark ) \cite{CMS-PAS-TOP-16-021}. 
 An accurate description of this observable is important in many searches for new physics phenomena and in measurements of the Higgs boson properties.~To improve the description of high jet multiplicities in \ttbar~events, the parameters  that are most sensitive to jet kinematics in \ttbar~events are selected and optimized.~The strong coupling parameter at $m_Z$ for initial-state radiation in the PS, \alfa, and the \hdamp parameter that controls the jet matching in the \PHEG+\PYTHIA8~\cite{Nason:2004rx,Frixione:2007vw,Alioli:2010xd} setup are tuned using Run 1 data on jet activity in \ttbar~events.~The Monash tune for \alfa adopts the $\alpha_s^{FSR}$ value (=0.1365) tuned to LEP event shapes. This is found to be the leading cause of overproduction of jets.~Using the jet multiplicity and leading additional jet $p_T$ distributions in the dilepton final state measured at $\sqrt{s}=8$ TeV \cite{Khachatryan:2015mva}, we tuned the \alfa and \hdamp parameters. In the fit, all other tune parameters are kept fixed to the ones in the \CO tune.~It is observed that \alfa impacts mostly $N_{jets}>3$, while \hdamp affects the ratio of 2-to-3-jet events and the leading additional jet $p_T$.~This is in agreement with the fact that the leading additional jet, in the \PHEG+\PYTHIA8~configuration, stems from the real radiation calculated by the \PHEG~ generator.~The tuning procedure yields \hdamp=$1.581^{+0.658}_{-0.585}\times m_t$ and \alfa=0.1108$^{+0.0145}_{-0.0142}$.~The tuned \alfa value agrees with the PDG value of $\alpha_s(M_Z)=0.1181\pm0.0011$ \cite{pdg} well within uncertainties. 
  
The probability for parton emission  is mainly constrained by the jet activity and the interplay between the hard and soft parts of the parton emissions.~However, it does not strongly constrain the global production of hadrons, i.e.~the underlying event (UE).~Therefore, \alfa constrained by \ttbar~jet kinematics can be used as a fixed input parameter in the UE tune.~See ref. \cite{CMS-PAS-TOP-16-021} for the details of the \CP~tune derived fixing \alfa to 0.1108.~It is found that both \PHEG+\PYTHIA8~and \MG\\ + \PYTHIA8 with FxFx merging \cite{Frederix:2012ps}  with tune \CP~describe the top quark data well (except for the top quark $p_T$ distribution, irrespective of the tune).~It is also observed that the global event variables $H_T$ or $S_T$ are not considerably  affected by the value of \alfa.~The comparison of different \ttbar~differential cross section predictions using \PHEG+\PYTHIA8~with the \CP~tune to the corresponding measurements at $\sqrt{s}=13$ TeV  yield an overall p-value of $<0.01$ when theory uncertainties are ignored. The p-value improves to 0.91 when theory uncertainties are included \cite{CMS:2017uda}. 

The \CP~tune is also used in a recent top quark mass measurement \cite{ref:top-17-007}, and the resulting top quark mass value is found to be consistent with the Run 1 results, based on a different MC generator and tune.~In addition, event tunes with alternative color-reconnection models (referred to in the previous section) are derived based on the \CP~event tune.~The top quark mass measured in bins of different kinematic variables with these different color-reconnection models show that there is no indication of a kinematic bias and that there is no significant sensitivity to tunes with different color-reconnection models. 

\section{Consistency of ME and PS Codes}
\begin{figure}[ht!]
\begin{center}
\subfloat[]{\includegraphics[width=0.4\textwidth]{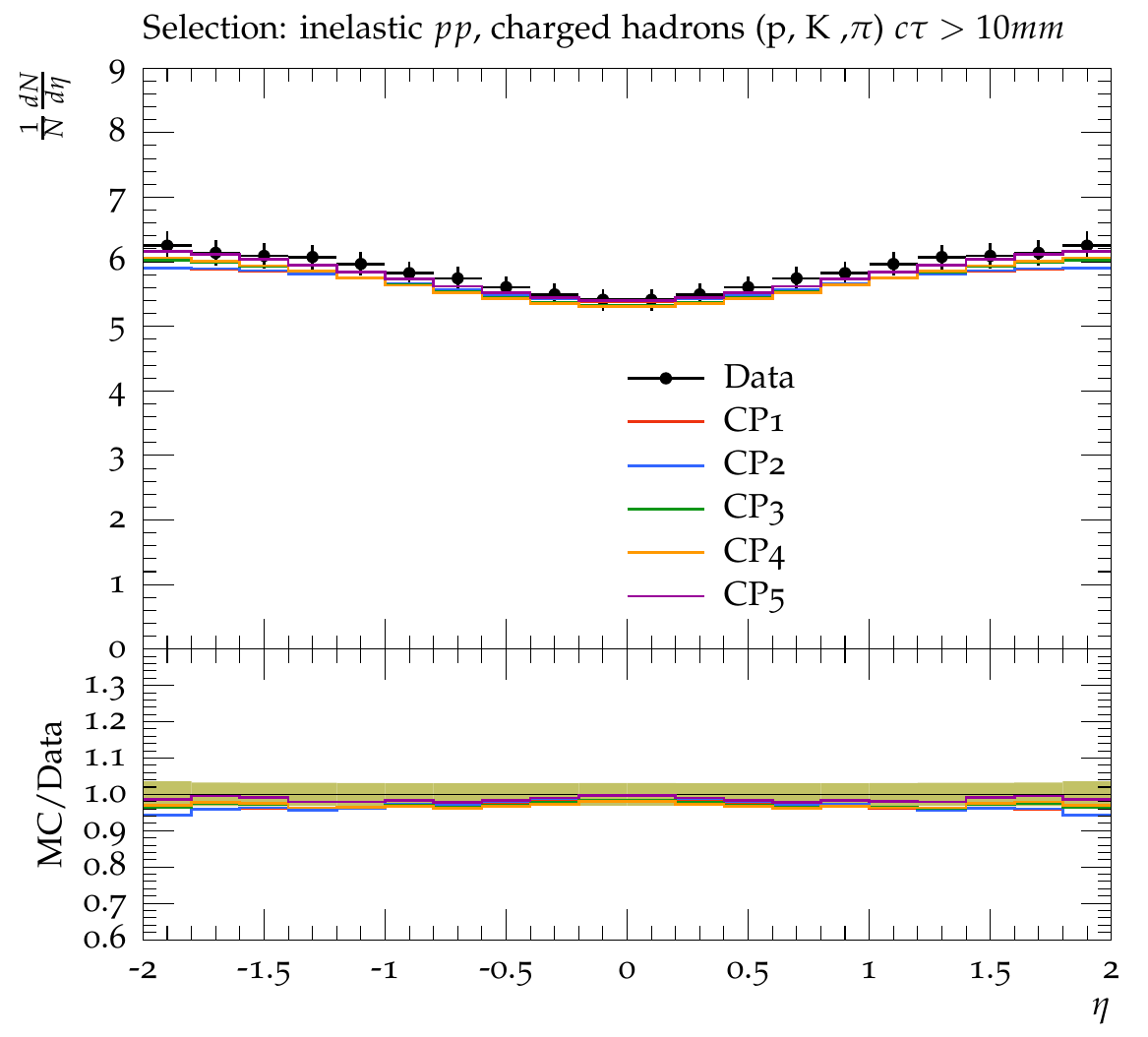}}
\subfloat[]{\includegraphics[width=0.4\textwidth]{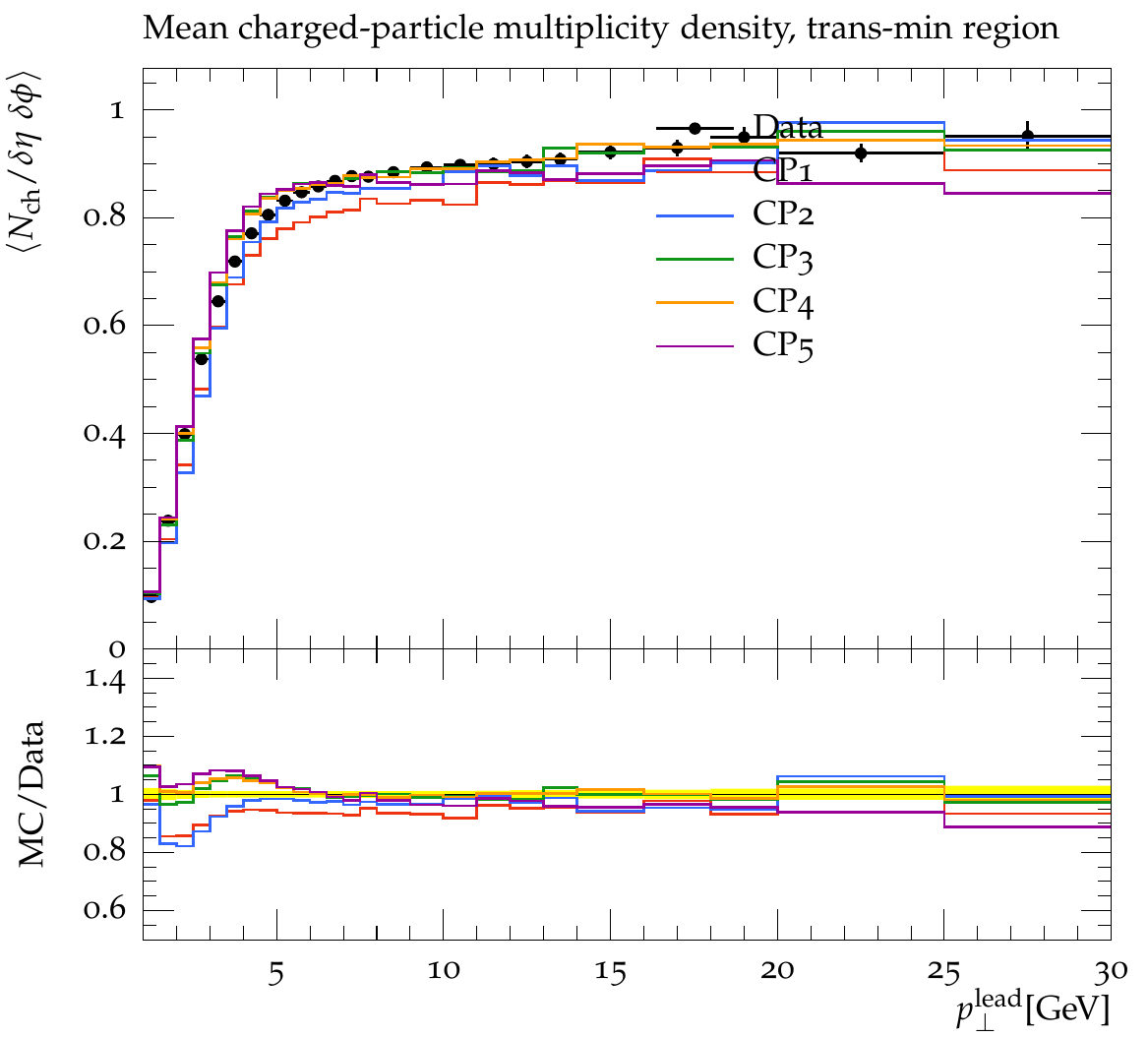}}
\caption{Predictions from tunes with NNPDF3.1 LO ($\alpha_s=0.130)$, NLO ($\alpha_s=0.118$), and NNLO ($\alpha_s=0.118$) tunes compared to CMS minimum bias (left)  and ATLAS UE (right) data \cite{pdf4lhcefe}. }
\label{fig:cp_comp}
\end{center}
\end{figure}

PDFs and $\alpha_s$ values are used in MC generators in several parts such as the hard partonic ME, in the computation of the ISR, as an input to the PS model, and to the MPI models.~The $\alpha_s$ values are typically  different at LO and NLO.~Traditionally, the order of the PDF is matched to the perturbative order of the ME calculation.
Using the same PDF set and the $\alpha_s$ value in the ME calculations and in the simulation of the various components of the PS is advocated in  \cite{cooperetal2011}, in particular, when the PS simulation is matched to higher-order matrix elements.~Different strategies are adopted; CMS and ATLAS tunes are traditionally based on LO PDFs, \PYTHIA8 tunes are mostly based on LO PDFs, new \SHERPA~tunes are based on NNLO PDFs, and \HERWIGalone7 provided tunes based on NLO PDFs.~We tested the effect of using different PDF orders of NNPDF sets in \PYTHIA8~among other parameter variations.~The CP1 and CP2 tunes use NNPDF3.1 LO ($\alpha_s=0.130)$, CP3 tune uses NNPDF3.1 NLO ($\alpha_s=0.118$), and CP4 and CP5 uses NNPDF3.1 NNLO ($\alpha_s=0.118$).~The predictions from these tunes are compared to data as shown in Figure \ref{fig:cp_comp}.~It is observed that UE and minimum bias data are described at the same level by tunes with LO, NLO, and NNLO NNPDF3.1 sets. 


\end{document}